# Planar Optical Nano-Antennas Resolve Cholesterol-Dependent Nanoscale Heterogeneities in the Plasma Membrane of Living Cells

Raju Regmi,[1,2] Pamina M. Winkler,[1] Valentin Flauraud,[3] Kyra J. E. Borgman,[1] Carlo Manzo,[1] Jürgen Brugger,[3] Hervé Rigneault,[2] Jérôme Wenger,[*,2] and María F. García-Parajo[*,1,4]

[1] *ICFO-Institut de Ciencies Fotoniques, The Barcelona Institute of Science and Technology, 08860 Barcelona, Spain,*
[2] *Aix Marseille Univ, CNRS, Centrale Marseille, Institut Fresnel, UMR 7249, Marseille, France,*
[3] *Microsystems Laboratory, Institute of Microengineering, Ecole Polytechnique Fédérale de Lausanne, 1015 Lausanne, Switzerland,*
[4] *ICREA, Pg. Lluís Companys 23, 08010 Barcelona, Spain*

E-mail: jerome.wenger@fresnel.fr; maria.garcia-parajo@icfo.eu

Abstract

Optical nano-antennas can efficiently confine light into nanoscopic hotspots enabling single-molecule detection sensitivity at biological relevant conditions. This innovative approach to breach the diffraction limit offers a versatile platform to investigate the dynamics of individual biomolecules in living cell membranes and their partitioning into cholesterol-dependent lipid nanodomains. Here, we present optical nano-antenna arrays with accessible surface hotspots to study the characteristic diffusion dynamics of phosphoethanolamine (PE) and sphingomyelin (SM) in the plasma membrane of living cells at the nanoscale. Fluorescence burst analysis and fluorescence correlation spectroscopy performed on nano-antennas of different gap sizes show that unlike PE, SM is transiently trapped in cholesterol-enriched nanodomains of 10 nm diameter with short characteristic times around 100 µs. Removal of cholesterol led to free diffusion of SM, consistent with the dispersion of nanodomains. Our results are consistent with the existence of highly transient and fluctuating nanoscale assemblies enriched by cholesterol and sphingolipids in living cell membranes, also known as lipid rafts. Quantitative data on sphingolipids partitioning into lipid rafts is crucial to understand the spatiotemporal heterogeneous organization of transient molecular complexes on the membrane of living cells at the nanoscale. The proposed technique is fully bio-compatible and thus provides various opportunities for biophysics and live cell research to reveal details that remain hidden in confocal diffraction-limited measurements.

The plasma membrane plays a major role in cell physiology and is thus of fundamental importance to living systems. The spatial organization and diffusion dynamics of its constituents (lipids and proteins) occurring at the nanoscale largely influence cellular processes such as transmembrane signaling, intracellular trafficking and cell adhesion.[1,2] Recent advances in cell biology have shown that the plasma membrane is significantly more complex than just a continuous fluidic system.[3–5] It has been postulated that sphingolipids, cholesterol and certain types of proteins can be enriched into dynamic nanoscale assemblies or nanodomains, also termed lipid rafts.[6–8] Lipid rafts have been defined as highly dynamic and fluctuating nanoscale assemblies of cholesterol and sphingolipids that in the presence of lipid- or protein-mediated activation events become stabi-lized to compartmentalize cellular processes.[2,5,9] However, the true nature of these nanodomains remains debated with many conflicting evidences and predicated domain sizes in the broad range of 10-200 nm, primarily because of their transient nature and nanoscopic sizes.[8–14]

Early investigations on membrane organization were mostly based on fluorescence recovery after photobleaching (FRAP)[15] and single particle tracking (SPT).[3,16] Both techniques are limited either in space (with $\mu m^2$ probe area in FRAP) or in time (with millisecond temporal resolution in SPT). Fluorescence correlation spectroscopy (FCS) is a widely adopted alternative for studying dynamics and biomolecular interactions.[17] FCS determines the average transit time from statistical averaging over many individual molecule diffusion events.[18] Its high temporal resolution together with its rather straightforward data analysis makes FCS an attractive tool to probe the spatiotemporal organization of cell membranes.[10,11,19] However, conventional FCS on confocal microscopes is unable to resolve the nanoscale organization of lipids due to the limited 200 nm spatial resolution set by diffraction.

Various approaches have been implemented over the past decade to breach the diffraction limit in FCS, but membrane studies have so far remained above a 40-50 nm detection size. Stimulated emission depletion microscopy (STED) constrains the excitation spot down to ~ 30 nm[20] and has been combined with FCS to explore the nanoscale dynamics occurring in lipid membranes on living cells.[21–24] An alternative strategy takes advantage of nanophotonic structures to engineer the light intensity distribution at the nanoscale.[25] Some notable designs include zero-mode waveguides,[26–31] bowtie structures,[32–34] gold nanorods[35] and sub-wavelength tip based NSOM probes.[36,37] These various approaches allow to confine the illumination light in the range of 50 to 100 nm. Resonant optical nanogap antennas have shown great potential to further constrain the laser light on a sub-20 nm scale[38] and greatly enhance the light-matter interactions.[39–42] However, so far the applications of such resonant nanogap antennas have been mostly employed to probe fluorescent molecules in solutions at high micromolar concentration. Recently, we have developed a new class of nano-antennas that maximizes access to the antenna hotspot region together with extreme planarity and biocompatibility.[43] This methodology has been validated using model lipid membranes,[44] underscoring its high potential to investigate the nanoscale architecture of living cell membranes.

In this work, we combine FCS with these planar optical nanogap antennas to investigate for the first time the nanoscopic organization of lipid rafts in the plasma membrane of living cells at a spatial resolution of 10 nm. The antenna design has been specifically developed for FCS with sub-diffraction spatial resolution.[41] It combines a central nanogap antenna to create the highly confined electromagnetic hotspot (of dimensions ~10 and 35 nm) surrounded by a rectangular cladding to prevent direct excitation of background molecules diffusing away from the central nanogap. By applying planarization, etch-back and template stripping methods, we have improved our initial design to produce arrays of nano-antennas with controlled gap sizes, sharp edges and planar hotspots facing the upper surface of the sample.[43] Using these planar nano-antennas with gap sizes down to 10 nm, we investigate here the diffusion dynamics of phosphoethanolamine (PE) and sphingomyelin (SM) on the plasma membrane of living Chinese hamster ovary (CHO) cells. Compared to earlier works using confocal FCS,[10,11,19] nanoaperture FCS,[26–30] or STED-FCS,[21–24] our study is the first to breach into the sub-30 nm spatial scale on living cell membranes. Together with cholesterol depletion experiments, we provide compelling evidence of short-lived cholesterol-induced ~ 10 nm nanodomain partitioning in plasma membranes and discuss the impact of these results in the context of lipid rafts.

The planar antenna platform contains multiple gold nano-antenna arrays with nominal gap sizes of 10 nm and 35 nm on which a circular cell culture well is mounted for live CHO cell culturing. Figure 1a,b depicts the strategy chosen for the fluorescence live cell experiment conducted on the nanogap antenna platform. A 640 nm laser light illuminates a single nano-antenna in the sample plane of an inverted microscope with a high-NA water immersion objective. Throughout this study, the linear polarization of the laser beam is set parallel to the antenna main axis so as to excite the nanogap mode.[43] A highly confined nanometric hotspot of illumination light is created on the surface of the nanogap region which is in direct contact with the adhered plasma membranes of living CHO cells. Importantly, the planarization strategy avoids possible curvature induced effects on the cell membrane and thus provides an ideal platform for live cell membrane research[29,30] (Supporting Information Fig. S1 shows AFM images indicating a planarity better than 3 nm for the top surface).

The cells were incubated on the nano-antennas at $37^o$ C for nearly 48 hours prior to the experiments to allow them to freely grow and adhere onto the antenna platform. Lipid analogs (either PE- or SM-BSA complexes) labeled with the lipophilic organic dye Atto647N were incorporated into the plasma membrane of the living cells just before the fluorescence measurements (see Methods for details on staining protocol). The choice of Atto647N as fluorescent dye allows an excellent overlap with the antenna's main plasmonic resonance (Fig. S2), maximizing the fluorescence enhancement in the nanogap. Figure 1c shows a representative confocal image of the morphology of the CHO cells adhered on a glass coverslip taken after the incorporation of the fluorescent lipid analogs.

Figure 2a,b shows representative single-molecule fluorescence time traces for PE and SM in the confocal and in the nano-antenna configuration. The resolution given by the diffraction limited spot in the confocal scheme does not allow to resolve heterogeneities that may occur at the sub-200 nm spatial scale, and as a result, the time traces for both PE and SM appear indistinguishable. In contrast, the highly confined surface hotspot originating from the 10 nm gap antenna clearly reveals differences in the characteristic diffusion dynamics for PE and SM. As shown in Fig. 2b, PE displays sharp peaks in the fluorescence time trace as a result of the sub-diffraction excitation hotspot created by the planar nanogap antenna. Unlike PE, the signature of SM is discernibly different at the nanoscale: the short bursts (a hallmark of free diffusion in ultra-small detection areas) are accompanied by high intensity bursts of significantly longer durations. This is a direct indication that the nanoscopic diffusion of SM on the cell membrane is deviating from free Brownian diffusion as compared to larger macroscopic scales.

To provide more quantitative information about the fluorescence time traces, we performed a fluorescence burst analysis to represent the distributions of burst duration versus burst intensity (see Methods).[34] Figure 2c shows the results for both PE and SM for the 10 nm nanogap antenna compared to the confocal configuration. The scatter plots for PE and SM in the confocal configuration show no visible differences with burst durations in the range 1-100 ms and intensities around 20-30 counts/ms. However, in stark contrast, the distributions obtained on the nano-antennas show clear differences between PE and SM. Diffusion events in sub-ms time scales are notably observed with the nano-antennas exhibiting burst durations as short as 10 µs. Such short events are more than two orders of magnitude faster than in the case of the confocal reference. Regarding the diffusion dynamics for PE (red dots) probed with the nanogap antennas a general trend can be deduced, namely, brighter events arise at shorter timescales. These can be understood as the detection of a "best burst event" directly resulting as a consequence of an individual molecule diffusing through the hotspot in the optimal position and orientation for maximum enhancement. The tighter the excitation beam confinement, the higher is the local intensity which leads to higher fluorescence intensity and shorter burst duration (see Supplementary Information Fig. S3 and S4 for additional fluorescent time traces and analysis on different antennas and cells). We thus relate the events with burst durations below 1 ms to the trajectories occurring within the nanogap region.[34] In the case of PE, the bursts with durations above 1 ms feature a lower intensity in the range of 20-70 counts/ms, which is only slightly increased as compared to the confocal level. We assign these longer burst duration events to the residual excitation of diffusing molecules within the larger 300 × 140 nm$^2$ box aperture region where the electromagnetic field intensity enhancement is negligible and thus comparable to the confocal reference.

In contrast to PE, SM probed with the nano-antenna arrays shows a significantly broader distribution of burst lengths against peak burst intensities (Fig. 2c). High intensities are observed for burst durations below and above 1 ms. Since these events were not observed for PE, we relate their occurrence to nanoscopic heterogeneities such as transient molecular complexes on the cell membrane hindering the diffusion of SM. To support this conclusion, we perturbed the cholesterol composition in the cell membrane with methyl-β-cyclodextrin (MCD) as cholesterol is expected to play a significant role in the formation and stability of the lipid nanodomains. The result of the burst analysis for SM after MCD treatment recovers a distribution which closely resembles the one for PE (Supporting Information Fig. S5). In other words, the intense bursts of duration between 0.1 and 10 ms disappear after cholesterol depletion, consistent with the loss of nanodomains. Altogether, the results from the fluorescence burst analysis demonstrate the benefits of planar nanogap antennas to explore the nanoscopic organization of lipids in live cell membranes. Clear differences between PE and SM diffusion dynamics are unveiled that otherwise would remain hidden in confocal measurements.

To further support these results, we performed fluorescence correlation spectroscopy (FCS) analysis. FCS records the fluorescence intensity fluctuations as the fluorophores transit through the detection spot. These fluctuations are analyzed by computing the temporal autocorrelation function, averaging over thousands of single-molecule diffusion events. We used two different gap sizes (10 and 35 nm) to quantify the lipid dynamics for increasing detection areas in cell membranes. Figure 3a,b shows the normalized correlation traces for PE and SM in case of the nano-antennas and the confocal reference. Each of these traces is taken on an individual nano-antenna (more traces are shown in Fig. S3 and S4 to demonstrate the consistency of our results). Similar to the burst analysis, we find no significant differences between the FCS curves for PE and SM for the confocal reference (gray circles in Fig. 3a,b the overlay of the confocal FCS data is shown in Fig. S6), yielding comparable diffusion times of 25 ± 4 ms (PE) and 30 ± 4 ms (SM), respectively. In the case of the nano-antennas, we observe that decreasing the gap size leads to a faster diffusion, confirming that the fluorescence signal stems from the nanogap region. We use a two-species model to fit the FCS data in order to account for the fluorescence contributions stemming from the nanogap and from the surrounding aperture area (see details in Methods section). A key feature in FCS is that the molecules contribute to the correlation amplitude in proportion to the square of their fluorescence brightness, hence the signal from molecules in the nanogap experiencing maximum enhancement will have a dominating contribution to the FCS curves.[45] The complete results and values for the FCS fits are detailed in the Supporting Information Tables S1-S3.

The differences between PE and SM diffusion dynamics are highlighted in Fig. 3c where a direct comparison of the FCS data for the 10 nm gap antenna is shown for different fluorescent lipid analogs. Contrarily to the confocal case (Fig. S6), the difference in diffusion times between the two lipids becomes more prominent at the nanoscale, with PE exhibiting diffusion times of 0.25 ± 0.06 ms and SM of 0.35 ± 0.04 ms. Moreover, after MCD treatment, the diffusion dynamics for cholesterol-depleted SM closely resembles that of PE with a diffusion time of 0.19 ± 0.03 ms (Fig. 3d). These FCS results confirm the presence of cholesterol-enriched nanodomains hindering the diffusion of SM, in agreement with the results found for the fluorescence burst analysis. In addition, we retrieved an anomaly value alpha for SM that depended on the probed area, deviating from unity as the illumination area reduced, from $\alpha \sim 0.85$ (for the 35 nm gap antenna) to $\alpha \sim 0.65$ (for the 10 nm gap antenna), which is fully consistent with hindered diffusion (see Table S1 to S3). In contrast, the $\alpha$ values were significantly larger and closer to unity for the cases of PE and SM after MCD treatment ($\alpha \sim 0.85$) and did not depend on the probe area, as expected for Brownian, unhindered diffusion.

To further analyze and exploit the FCS data we take advantage of the large number of planar nano-antennas with controlled gaps to carry out a FCS analysis over 60 different antennas and cells. This approach follows the so-called FCS diffusion law,[19,28] which is a representation of the diffusion time versus the detection area. Extrapolation of the experimental curve to the intercept with the time axis provides information on the type of diffusion exhibited by the molecule, i.e., free diffusion is characterized by a linear curve crossing the origin (0,0), while hindered diffusion due to the occurrence of nanodomains leads to a positive intercept on the time axis.[19,28] The nano-antenna detection area was estimated as the product of the gap size (measured by transmission electron microscopy TEM) times the full width at half maximum for the intensity profile along the direction perpendicular to the antenna main axis (simulated by finite difference time domain (FDTD) method, see Supporting Information Fig. S7 for simulations results). Moreover, the area sizes were further confirmed by calibration measurements both on freely diffusing dyes in solution and on pure PE lipid bilayers using antennas of different gap sizes.[44] The 10 and 35 nm gap antennas correspond respectively to 300 nm$^2$ and 1250 nm$^2$ illumination areas. As the diffusion time proportionally scales with the detection area, the diffusion coefficient $D$ is retrieved from the slope of the linear fit matching the measured transient diffusion times obtained from the FCS curves versus the effective detection areas according to the relation $D = \text{probe area}/4 \times \tau_{\text{diff}}$.[18]

Figure 4a-c summarizes the characteristic diffusion times for PE, SM and SM after cholesterol depletion for two antenna gap areas. The extension to include the confocal data is shown in Fig. S8. From these graphs we derive the following three values plotted in Fig. 4d-f: the diffusion coefficient (from the slope), the time axis intercept (by extrapolating the linear fit for vanishing probe area) and the normalized spread in the data points (defined as the width of upper and lower quartiles divided by the median value). The diffusion coefficients derived from nano-antenna measurements are $D_{\text{PE}} = 0.44 \pm 0.07$ µm$^2$/s, $D_{\text{SM}} = 0.38 \pm 0.19$ µm$^2$/s and $D_{\text{MCD-SM}} = 0.46 \pm 0.07$ µm$^2$/s (Fig. 4d) and they are consistent with the confocal measurements and values reported independently using STED-FCS.[21] These coefficients represent the diffusion speed in the lipidic region between the nanodomains, with an additional contribution from diffusion within the nanodomains and diffusion of the domains themselves.

Extrapolating the fits in Fig. 4a-c towards diminishing probe area leads to the intercepts with the time axis as summarized in Fig. 4e. The almost zero intercept hitting the origin observed for PE confirms the expected free Brownian motion diffusion mode. In stark contrast, SM features a positive y-intercept of about 100 µs, which highlights a significant deviation from free Brownian diffusion and the occurrence of nanoscopic domains hindering SM diffusion. Depletion of cholesterol results on SM diffusion with a close-to-zero time intercept, demonstrating the crucial role of cholesterol establishing the nanodomains and hindering SM diffusion. Such small nanoscale heterogeneities have never been detected so far with confocal microscopy, although STED-FCS down to 1000 nm$^2$ detection area could infer their occurrence.[21] Our results are fully aligned with these previous findings and importantly, we further reduce the detection areas down to 300 nm$^2$.

Lastly, we take a closer look at the statistical dispersion of the diffusion times for each gap area, and introduce the normalized data spread as the width from upper to lower quartiles divided by the median value (Fig. 4f). The spread in diffusion times for PE and SM after MCD treatment remains under 25% and can be partially assigned to nanometer variations of the gap size between nanoantennas.[44] These variations stem from the nanofabrication process as a consequence of the finite grain size of gold and/or the scattering of electrons used during the electron beam lithography. In contrast to PE and MCD-SM, the data for SM features a significantly higher statistical dispersion around 50%, which cannot be related solely to dispersion in the nanoantenna sample, but instead it results from large variations in the SM diffusion behavior, as already noted for the fluorescence burst analysis (Fig. 2c). These results are fully consistent with the presence of cholesterol-enriched nanodomains affecting SM diffusion.

Altogether, our results provide compelling evidence for the existence of highly transient and fluctuating nanoscale assemblies of sterol and sphingolipids in living cell membranes. These experimental observations stand in excellent agreement with the notion that without stabilizing proteins, lipid rafts can be viewed as intrinsic nanoscale membrane heterogeneities that are small and highly transient.[2,6–8] We estimate the characteristic residence time of the fluorescent SM lipid analogs in the nanodomain from the y-intercept in Fig. 4b,e, and find a value around 100 µs . The typical size of the nanodomains could in principle also be deduced from the FCS diffusion laws which should feature a characteristic transition from confined to normal diffusion.[19,28] As we do not observe this characteristic transition in our data, we conclude that the typical size of the nanodomain is smaller than the smallest gap size of our nanoantenna, that is 10 nm. Both the typical nanodomain size about 10 nm and the transient time about 100 µs stand in good agreement with the predictions from stochastic models[46] and recent high-speed interferometric scattering (iSCAT) measurements on mimetic lipid bilayers containing cholesterol.[12] We believe that this shorter characteristic time as compared to earlier experimental works using STED-FCS[21–23] is related to the smaller 10 nm resolution achieved in our case. The nanoantenna approach is straightforward to implement on any confocal microscope equipped for FCS as contrarily to STED, it does not require adding any supplementary illumination beam. As additional advantage, the excellent planarity of the surface rules out any artefact potentially induced by the curvature of the cell membrane.[29] We believe that these advantages and the excellent spatiotemporal resolution largely compensate for the need for nanofabrication and the more complex FCS fitting procedure.

In conclusion, we have demonstrated the promising approach of exploiting planar optical nano-antennas with accessible surface nanogaps to investigate the nanoscale architecture of live cell membranes. The key strengths of our approach rely on the 10 nm spatial resolution combined with a microsecond time resolution on a nearly perfectly flat substrate compatible with live cell culturing. The single-molecule data on nanoantennas reveal striking differences between PE and SM diffusion dynamics that remain hidden in confocal measurements. Fluorescence burst and correlation spectroscopy analysis for PE are consistent with a free Brownian diffusion model. In contrast, the diffusion dynamics of SM at the nanoscale show heterogeneities in both time and space which are cholesterol dependent. Indeed, removal of cholesterol leads to a recovery of free Brownian diffusion for SM, consistent with the loss of nanodomains. Our results are consistent with the existence of dynamic nanodomains on the plasma membranes of living cells of ~10 nm diameter which is comparable to our measurement gap size. The corresponding transient trapping times are short of about ~100 µs. We believe that the combination of optical nano-antennas with fluorescence microscopy has a high potential to investigate the dynamics and interactions of raft associated proteins and their recruitment into molecular complexes on the plasma membrane of living cells. The proposed technique is fully bio-compatible and thus provides ample opportunities for biophysics and live cell research with single-molecule sensitivity at nanometric and (sub)microsecond spatiotemporal resolution, far beyond the diffraction limit of light.

## Methods

Planar Nanogap Antenna Array Fabrication Large scale nano-antenna arrays with surface nanogaps were fabricated by combining electron beam lithography with planarization, etch back and template stripping.[43] First, EBL was used to pattern features with negative tone hydrogen silsesquioxane (HSQ) resist on top of 100 mm silicon wafer. A 50 nm thick gold film was then deposited by electron beam evaporation at low temperature to reduce the gold grain size. Flowable oxide (Dow Corning FOX-16) was spun to planarize the overall structure and was followed by an etch back step to selectively remove the sacrificial top metal layer clearing the aperture geometry. A 30 s etch with hydrofluoric acid diluted 1:10 in deionized water was then used to clear out the residual HSQ in the antennas. Finally, the antenna structures were embedded in an UV curable adhesive polymer (OrmoComp, Microresiste Technology GmbH) and stripped away from the silicon wafers. The narrowest gap region lying at the bottom of the structure due to the metal diffusion during the evaporation, was now flipped over to maximize the contact with the sample providing direct accessibility to the plasmonic antenna hotspot. This method is fully scalable and shows excellent geometry control and planarity (see Supporting Information Fig. S1).

Cell culture, Atto647N-labeling and Cholesterol depletion of CHO cells CHO cells were seeded on a coverslip containing planar nano-antennas with surface nanogaps and were allowed to grow and spontaneously attach at $37^o$ C in a controlled atmosphere with 5% of $CO_2$ for nearly 48 hours. Lipid conjugates were separately prepared by labeling 1,2-Dipalmitoyl-sn-glycero-3-phosphoethanolamine (DPPE) and Sphingomyelin (SM) with the organic dye Atto647N (from Invitrogen) as described in Ref.[21] Prior to the fluorescence experiments, the lipid analogues were incorporated in the cell membrane during a 3 mins incubation period at room temperature dissolved in the corresponding medium for CHO cells (Ham's F12 nutrient mixture). Stained cell cultures were rinsed and washed to remove residual dye molecules before placing the sample coverslip on the piezo-stage of an inverted microscope to carry out the measurements. For cholesterol depletion experiments, the CHO cells were incubated in serum free buffer with 10 mM methyl-$\beta$-cyclodextrin (MCD) for 30 mins at $37^o$C, and then the fluorescent labeling was carried out as previously described. All fluorescence stainings were performed at a ~300 nM concentration of Atto647N and the measurements were completed within 30 mins after the incorporation of the fluorescent analogs. From the number of detected fluorescence bursts (Fig. 2c) and the FCS amplitude, we estimate that the density of fluorescent lipids for the antenna experiments is on the order of 20 to 80 probes per $\mu m^2$.

Experimental Setup  The experiments were performed with a commercial MicroTime 200 setup equipped with an inverted confocal microscope (Olympus 60×, 1.2 NA water-immersion objective) and a three-axis piezoelectric stage (PhysikInstrumente, Germany) allowing to select individual nano-antennas. A linearly polarized 640 nm picosecond laser diode (Pico-Quant LDH-D-C-640) in continuous wave mode was used to resonantly excite individual nano-antennas with the laser linear polarization aligned with the main axis of the antenna dimer. The emitted fluorescence signal was collected in epi-detection mode through a dichroic mirror and the signal was finally split into two avalanche photodiodes (PicoQuant MPD-50CT). An emission filter and a band pass 650-690 nm filter just before each detector was used to eliminate the scattered light by the excitation laser. A 30 µm pinhole in the detection arm yielded 0.5 fl confocal detection volume at the sample plane. The fluorescence time traces were recorded on a fast time-correlated single photon counting module in the time-tagged time-resolved mode (PicoQuant MPD-50CT). All the fluorescence measurements were performed by illuminating the sample at an excitation power density of ∼ 2-3 kW/cm$^2$. The measurements were acquired for a typical run time of 50 s and the correlation amplitudes were computed for ∼ 20 s windows with the commercial software package SymPhoTime 64. Cells were cultured on different antenna samples, each sample containing different gap sizes.

Fluorescence Burst Analysis  Single-molecule fluorescence time traces were acquired in the Tagged Time-Resolved (TTTR) mode (recording each event at its arrival time) with 4 ps temporal resolution. Fluorescence bursts analysis was then carried out with a likelihood-based algorithm to test the null hypothesis (no burst, recording compatible with background noise) against the hypothesis that a single-molecule burst arises as a consequence of a molecule crossing the excitation area. Probabilities associated to false positive and missing event errors were both set to 10$^{-3}$.[49]

Fluorescence Correlation Spectroscopy  The temporal fluctuations of the fluorescence intensity $F(t)$ around the average value were analyzed to compute the temporal correlation $G(\tau) =$ $\langle \delta F(t).\delta F(t+\tau)\rangle/\langle F(t)\rangle^2$, where $\delta F(t) = F(t) - \langle F(t)\rangle$ is the fluctuation of the fluorescence signal arising due to the molecules crossing the detection volume mediated by Brownian diffusion, $\tau$ is the delay (lag) time, and $\langle\rangle$ indicates time averaging. The mobility of molecules shows strong dependence on the local environment and thus in living systems the concept of ideal Brownian diffusion may not always hold true. Considering possible anomalous diffusion in living cells, the temporal correlation of the fluorescence intensity $F$ can be written as:[18]

$$G(\tau) = \sum_{i=1}^{n_{\text{diff}}} \frac{\rho(i)}{1 + \left[\frac{\tau}{\tau_{\text{diff}}(i)}\right]^{\alpha(i)}}$$

where $\tau_{diff}(i)$ the average residence time of the $i^{th}$ diffusing modality, $\rho(i)$ denotes the respective amplitude contribution and $\alpha(i)$ being anomaly parameter of the same.[21] We find that the FCS curves recorded with a nanoscopic illumination can only be fitted with a model assuming two different diffusion modalities (*i.e.* $n_{\text{diff}} = 2$).

To define the probe areas used in the FCS diffusion laws (Fig. 4a-c), we use the product of the gap size (measured by TEM) by the full width at half maximum for the intensity profile along the direction perpendicular to the antenna main axis (computed by FDTD), following a calibration for model lipid membranes.[44] Therefore, 10 nm and 35 nm gap sizes are associated respectively to 300 and 1250 nm$^2$ probe areas.

## Supporting Information

Supporting Information available: AFM image of antenna array, overlap between antenna's resonance and fluorescence spectra, fluorescence data for PE on different nano-antennas, overlay of FCS curves from different nanoantennas, representative time trace of cholesterol depleted SM, overlay of the confocal FCS data for PE and SM, FDTD simulations of intensity distributions, FCS diffusion laws with confocal data, fitting parameters for FCS curves.

# Additional information

CM present address: Universitat de Vic, Universitat Central de Catalunya (UVic-UCC), C. de la Laura 13, 08500 Vic, Spain

The authors declare no competing financial interests.

# Acknowledgement

The authors thank Merche Rivas Jiménez, Felix Campelo and Erik Garbacik for technical support and fruitful discussions. The research leading to these results has received funding from the European Commission's Seventh Framework Programme (FP7-ICT-2011-7) under grant agreements ERC StG 278242 (ExtendFRET) and 288263 (NanoVista). Financial support by the Spanish Ministry of Economy and Competitiveness ("Severo Ochoa" Programme for Centres of Excellence in R&D (SEV-2015-0522) and FIS2014-56107-R grants) and Fundacion Privada Cellex is gratefully acknowledged. RR is supported by the Erasmus Mundus Doctorate Program Europhotonics (Grant 159224-1-2009-1-FR-ERA MUNDUS-EMJD). PMW is supported by the ICFOstepstone Fellowship, a COFUND Doctoral Program of the Marie Skłodowska-Curie Action from the European Commission. CM acknowledges funding from the Spanish Ministry of Economy and Competitiveness and the European Social Fund (ESF) through the Ramón y Cajal program 2015 (RYC-2015-17896).

# References

(1) Brown, D. A.; London, E. *J. Biol. Chem.*, 2000, *275*, 17221-17224.

(2) Lingwood, D.; Simons, K. *Science*, 2010, *327*, 46-50.

(3) Kusumi, A.; Nakada, C.; Ritchie, K.; Murase, K.; Suzuki, K.; Murakoshi, H.; Kasai, R. S.; Kondo, J.; Fujiwara, T.; *Annu. Rev. Biophys. Biomol. Struct.*, 2005, *34*, 351-378.

(4) Gowrishankar, K.; Ghosh, S.; Saha, S. C. R.; Mayor, S.; Rao, M. *Cell* 2012, *149*, 1353-1367.

(5) Sezgin, E.; Levental, I.; Mayor, S.; Eggeling, C. *Nat. Rev. Mol. Cell Biol.*, 2017, *18*, 361-374.

(6) Mayor, S.; Rao, M. *Traffic*, 2004, *5*, 231-240.

(7) Simons, K.; Gerl, M. J. *Nat. Rev. Mol. Cell Biol.*, 2010, *11*, 688-699.

(8) Hancock, J. F. *Nat. Rev. Mol. Cell Biol.*, 2006, *7* 456-62.

(9) Pike, L. J. *J. Lipid Res.* 2006, *47*, 1597-1598.

(10) Lenne, P. F.; Wawrezinieck, L.; Conchonaud, F.; Wurtz, O.; Boned, A.; Guo, X. J.; Rigneault, H.; He, H. T.; Marguet, D. *EMBO J.* 2006, *25*, 3245-3256.

(11) Marguet, D.; Lenne, P. F.; Rigneault, H.; He, H. T. *EMBO J.* 2006, *25*, 3446-3457.

(12) Wu, H.-M.; Lin, Y.-H.; Yen, T.-C.; Hsieh, C.-L. *Sci. Rep.* 2016, *6*, 20542.

(13) van Zanten, T. S.; Cambi, A.; Koopman, M.; Joosten, B.; Figdor, C. G.; Garcia-Parajo, M. F. *Proc. Natl. Acad. Sci. USA.*, 2009, *106*, 18557-18562.

(14) van Zanten, T. S.; Gómez, J.; Manzo, C.; Cambi, A.; Buceta, J.; Reigada, R.; Garcia-Parajo, M. F. *Proc. Natl. Acad. Sci. USA.*, 2010, *107*, 15437-15442.

(15) Meder, D. M.; Moreno, J.; Verkade, P.; Vaz, W. L. C.; Simons, K. *Proc. Natl. Acad. Sci. USA.*, 2006, *103*, 329-334.

(16) Dietrich, C.; Yang, B.; Fujiwara, T.; Kusumi, A.; Jacobson, K. *Biophys. J.*, 2002, *82*, 274-284.

(17) Maiti, S.; Haupts, U.; Webb, W. W. *Proc. Natl. Acad. Sci. USA.*, 1997, *94*, 11753-11757.

(18) Bacia, K.; Kim, S. A.; Schwille, P. *Nat. Methods*, 2006, *3*, 83-89.

(19) Wawrezinieck, L.; Rigneault, H.; Marguet, D.; Lenne, P.-F. *Biophys. J.*, 2005, *89*, 4029-4042.

(20) Kastrup, L.; Blom, H.; Eggeling, C.; Hell, S.W. *Phys. Rev. Lett.*, 2005, *94(17)*, 1-4.

(21) Eggeling, C.; Ringemann, C.; Medda, R.; Schwarzmann, G.; Sandhoff, K.; Polyakova, S.; Belov, V. N.; Hein, B.; von Middendorff, C.; Schönle, A.; Hell, S. W. *Nature*, 2009, *457*, 1159-1162.

(22) Mueller, V.; Ringemann, C.; Honigmann, A. ; Schwarzmann, G.; Medda, R.; Leutenegger, M.; Polyakova, S.; Belov, V. N.; Hell, S. W.; Eggeling, C. *Biophys. J.*, 2011, *101*, 1651-1660.

(23) Honigmann, A.; Mueller, V.; Ta, H.; Schoenle, A.; Sezgin, E.; Hell, S. W.; Eggeling, C. *Nat. Commun.* 2014, *5*, 5412.

(24) Vicidomini, G.; Ta, H.; Honigmann, A. ; Mueller, V.; Clausen, M. P.; Waithe, D.; Galiani, S.; Sezgin, E.; Diaspro, A.; Hell, S. W.; Eggeling, C. *Nano Lett.*, 2015, *15*, 5912-5918.

(25) Punj, D.; Ghenuche, P.; Moparthi, S. B.; de Torres, J.; Grigoriev, V.; Rigneault, H.; Wenger, J. *WIREs Nanomed. Nanobiotechnol.*, 2014, *6*, 268-282.

(26) Edel, J. B.; Wu, M.; Baird, B.; Craighead, H. G. *Biophys. J.* 2005, *88*, L43-L45.

(27) Moran-Mirabal, J. M.; Torres, A. J.; Samiee, K. T.; Baird, B. A.; Craighead, H. G. *Nanotechnology* 2007, *18*, 195101.

(28) Wenger, J.; Conchonaud, F.; Dintinger, J. Wawrezinieck, L.; Ebbesen, T. W.; Rigneault, H.; Marguet, D.; Lenne, P.-F. *Biophys. J.*, 2007, *92*, 913-919.

(29) Kelly, C. V.; Baird, B. A.; Craighead, H. G. *Biophys. J.* 2011, *100*, 34-36.

(30) Kelly, C. V.; Wakefield, D. L.; Holowka, D. A.; Craighead, H. G.; Baird, B. A. *ACS Nano* 2014, *8*, 7392-7404.

(31) Richards, C. I.; Luong, K.; Srinivasan, R.; Turner, S. W.; Dougherty, D. A.; Korlach, J.; Lester, H. A. *Nano Lett.* 2012, *12*, 3690-3694.

(32) Lohmüller, T.; Iversen, L.; Schmidt, M.; Rhodes, C.; Tu, H. L.; Lin, W. C.; Groves, J. T. *Nano Lett.* 2012, *12*, 1717-1721.

(33) Mivelle, M.;, Van Zanten, T. S; Neumann, L.; van Hulst, N.; García-Parajo, M. F. *Nano Lett.*, 2012, *12 (11)*, 5972-5978.

(34) Flauraud, V.; van Zanten, T.S.; Mivelle, M.; Manzo, C.; García-Parajo, M.F.; Brugger, J. *Nano Lett.* 2015, *15*, 4176-4182.

(35) Pradhan, B.; Khatua, S.; Gupta, A.; Aartsma, T.; Canters, G.; Orrit, M. *J. Phys. Chem. C* 2016, *120*, 25996-26003.

(36) García-Parajo, M. F. *Nat. Photonics*, 2008, *2*, 201-203.

(37) Manzo, C.; Van Zanten, T. S; García-Parajo, M. F. *Biophys. J.*, 2011, *1000*, 08-10.

(38) Novotny, L.; van Hulst, N. *Nat. Photonics* 2011, *5*, 83-90.

(39) Kinkhabwala, A.; Yu, Z. F.; Fan, S. H.; Avlasevich, Y.; Mullen, K.; Moerner, W. E. *Nat. Photonics* 2009, *3*, 654-657.

(40) Acuna, G. P.; Möller, F. M.; Holzmeister, P.; Beater, S.; Lalkens, B.; Tinnefeld, P. *Science* 2012, *338*, 506-510.

(41) Punj, D.; Mivelle, M.; Moparthi, S.B.; van Zanten, T.S.; Rigneault, H.; van Hulst, N.F.; García-Parajo, M.F.; Wenger, J. *Nat. Nanotechnol.* 2013, *8*, 512-516.

(42) Puchkova, A.; Vietz, C.; Pibiri, E.; Wünsch, B.; Sanz Paz, M.; Acuna, G. P.; Tinnefeld, P. *Nano Lett.* 2015, *15*, 8354-8359.

(43) Flauraud, V.; Regmi, R.; Winkler, P. M.; Alexander, D. T. L.; Rigneault, H.; van Hulst, N. F.; García-Parajo, M. F.; Wenger, J.; Brugger, J. *Nano Lett.* 2017, *17*, 1703-1710.

(44) Winkler, P. M.; Regmi, R.; Flauraud, V.; Brugger, J.; Rigneault, H.; Wenger, J.; García-Parajo, M. F. *ACS Nano* 2017, *11*, 7241-7250.

(45) Langguth, L.; Koenderink, A. F. *Opt. Express* 2014, *22*, 15397-15409.

(46) Nicolau, D. V.; Burrage, K.; Parton, R. G.; Hancock, J. F. *Mol. Cell. Biol.* 2006, *26*, 313-323.

(47) Spillane, K. M.; Ortega-Arroyo, J.; de Wit, G.; Eggeling, C.; Ewers, H.; Wallace, M. I.; Kukura, P. *Nano Lett.* 2014, *14*, 5390-5397.

(48) Spindler, S.; Ehrig, J.; König, K.; Nowak, T.; Piliarik, M.; Stein, H. E.; Taylor, R. W.; Garanger, E.; Lecommandoux, S.; Alves, I. D.; Sandoghdar, V. *J. Phys. Appl. Phys.* 2016, *49*, 274002.

(49) Zhang, K.; Yang, H. *J. Phys. Chem. B*, 2005, *109*, 21930-21937.

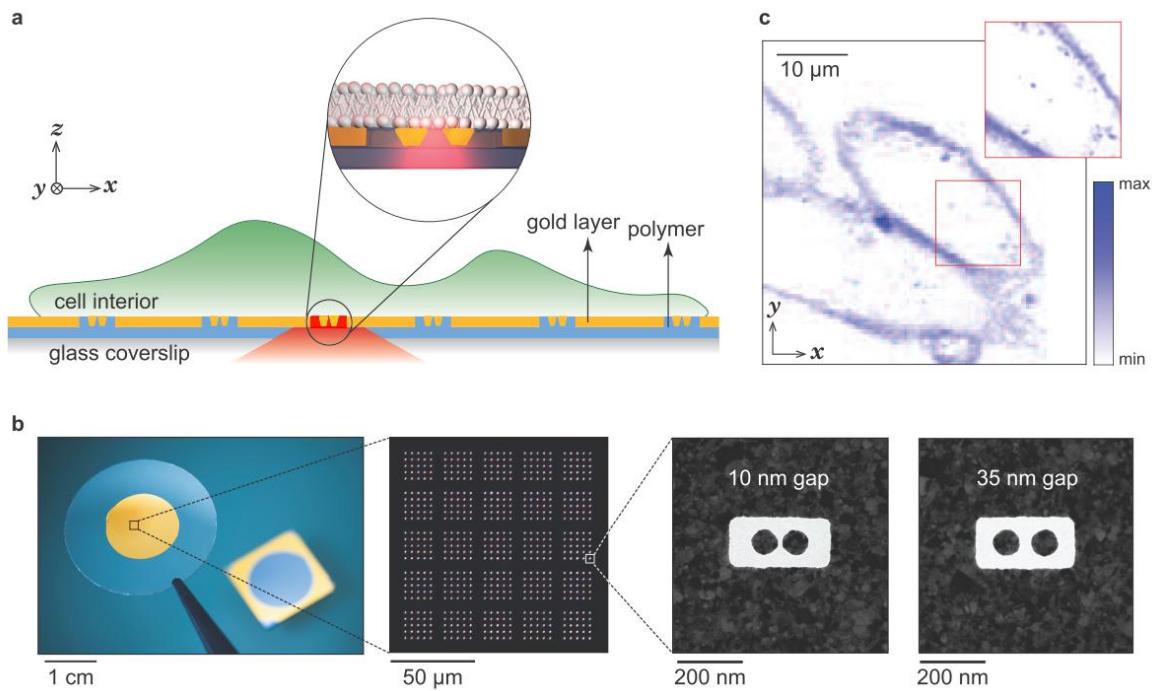

Figure 1: Planar gold nano-antenna arrays for probing single-molecule dynamics in the plasma membrane of living cells. (a) CHO cells are seeded onto a microscopic coverslip containing multiple planar nano-antennas with 10 and 35 nm gap sizes. The inset shows the cross section of the antenna-in-box stripped and embedded into a polymer, bringing the region of maximum electromagnetic field intensity onto the surface in direct contact with the plasma membrane of living cells. (b) From left to right: macro-photograph of a coverslip with a stripped Au film with large-scale planar antenna arrays; dark field optical micrograph of a small portion of the antenna arrays showing here 625 antennas with 10 nm nominal gap size; transmission electron microscope (TEM) images of antennas with 10 and 35 nm gap size. (c) Confocal image of CHO cells showing the morphology after incorporating the fluorescent SM lipid analog labeled with Atto647N.

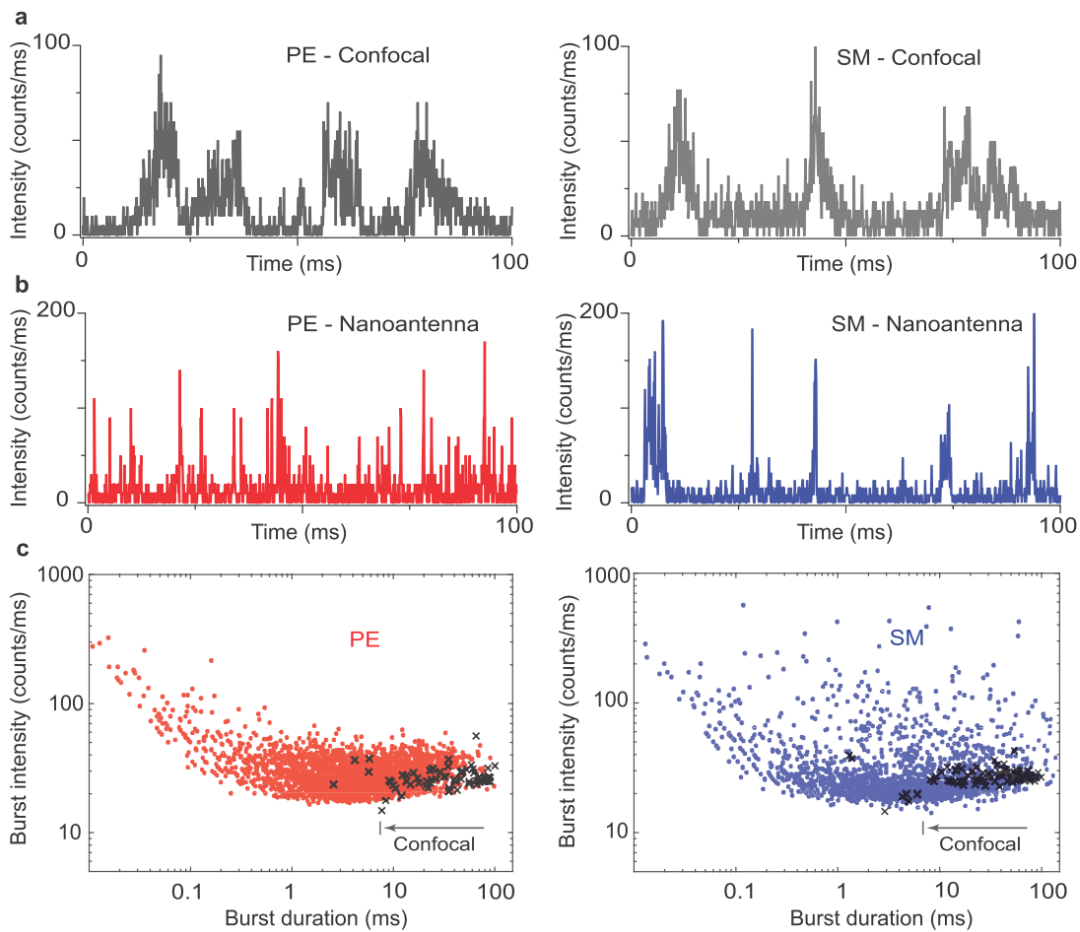

Figure 2: Single-molecule fluorescence time traces in living CHO cells. (a,b) Fluorescence time traces for phosphoethanolamine (PE, left) and sphingomyelin (SM, right) labeled with Atto647N recorded with confocal (a) and with a 10 nm gap planar nano-antenna (b). The binning time is 0.1 ms for all traces. The diffraction-limited spot in the confocal configuration cannot resolve the nanoscopic and heterogeneous membrane organization, thus results in indistinguishable fluorescence time traces for both PE and SM. However, the highly confined nano-antenna hotspot reveals clear differences in the diffusion dynamics of PE and SM. (c) The fluorescence time traces are analyzed to produce scatter plots showing the distribution of fluorescence burst intensity versus burst duration. Single-molecule events in sub-ms time scales are observed with nano-antennas (color dots) as the confined electromagnetic hotspots allow to probe the dynamics occurring beyond the diffraction limit. Single molecule events obtained with confocal illumination are shown for comparison (black dots).

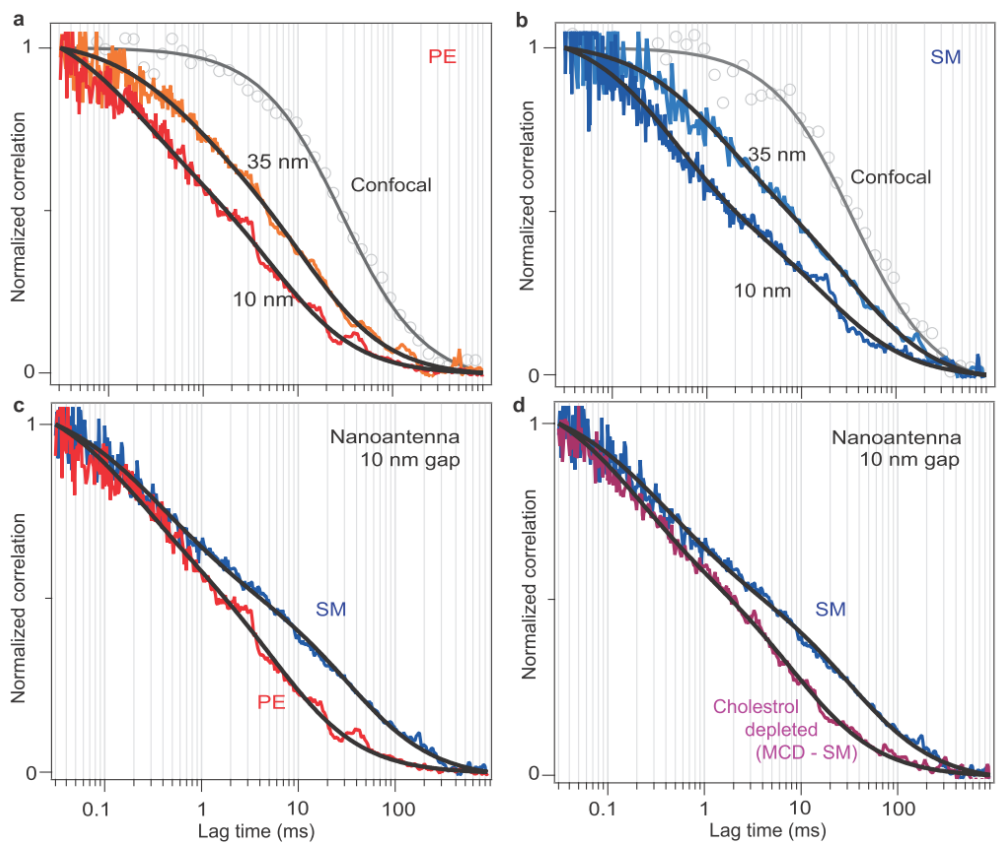

Figure 3: Nano-antenna FCS on living cell membranes. (a,b) Normalized fluorescence correlation curves for Atto647N labeled PE (a) and SM (b) lipid analogs probed with nano-antennas of varying gap size. The color lines are experimental data and the black curves are numerical fits. Each FCS trace is a representative example taken on an individual nano-antenna. FCS curves recorded on different nano-antennas and different cells are shown in Fig. S4. The diffraction-limited confocal measurements are shown in gray for direct comparison. (c) Comparison of FCS curves for PE and SM for a 10 nm gap antenna. Unlike the confocal reference, the nano-antenna reveals clear differences between the dynamics of PE and SM at the nanoscale. (d) After cholesterol depletion, the SM diffusion dynamics are significantly faster and resemble the PE case.

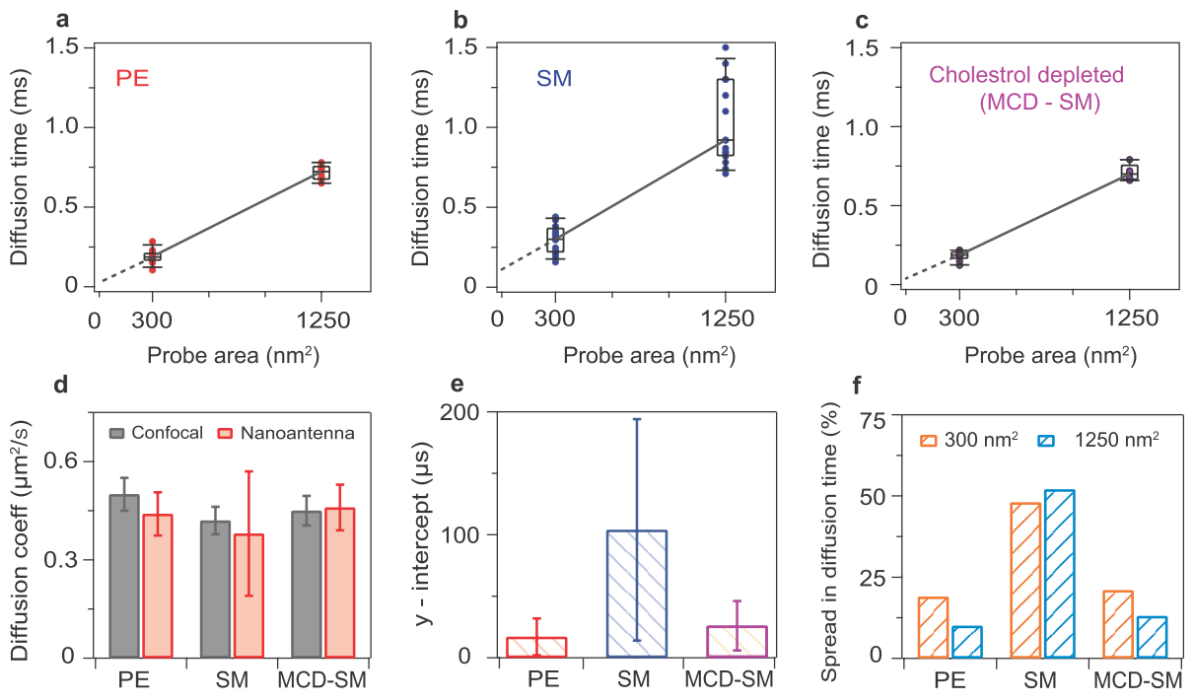

Figure 4: Characteristic diffusion dynamics of membrane lipids probed with ultra-confined nano-antenna hotspots. The diffusion time measured by FCS (for 60 different nano-antennas) is plotted as a function of the probe area for PE (a), SM (b), and SM after MCD treatment(c). The solid lines are linear fits through the median values. In the case of free diffusion, the origin (0, 0) is aligned with the expected line, while a positive intercept at the y-axis denotes hindered diffusion due to nanodomains. (d) The diffusion coefficients computed from the slopes in a-c are compared with confocal results. (e) y-axis intercept deduced from the linear fits in a-c. PE and MCD-treated SM show near-zero y-intercept consistent with free diffusion, while the significant y-intercept for SM indicates that the diffusion is constrained by nanodomains. (f) Normalized spread in diffusion time (width of upper and lower quartiles / median) in each case. The large dispersion observed is SM is another indication that sphingolipids are preferentially recruited into transient nanoscopic domains.

23